\documentclass[article, twocolumn, aps, prl, groupedaddress, showpacs, superscriptaddress, nofootinbib]{revtex4-1}

\usepackage{amsmath}
\usepackage{amssymb}
\usepackage{txfonts}
\usepackage{mathrsfs} 
\usepackage{gensymb}
\usepackage{graphicx,bm,units,yfonts}
\usepackage[bottom]{footmisc}
\usepackage{perpage}
\MakePerPage{footnote}

\usepackage[table]{xcolor}

\usepackage{hyperref}

\usepackage{cancel}

\usepackage[T1]{fontenc}
\usepackage{lmodern}

\begin{document}

\title{Observation of Flat Frequency Bands at Open Edges and Antiphase Boundary Seams in Topological Mechanical Metamaterials}

\author{Kai Qian}
\affiliation{Department of Physics, New Jersey Institute of Technology, Newark, New Jersey, USA}

\author{Linghua Zhu}
\affiliation{Department of Physics, Virginia Tech, Blacksburg, Virginia, USA}

\author{Keun Hyuk Ahn}
\email{kenahn@njit.edu}
\affiliation{Department of Physics, New Jersey Institute of Technology, Newark, New Jersey, USA}

\author{Camelia Prodan}
\email{cprodan@njit.edu}
\affiliation{Department of Physics, New Jersey Institute of Technology, Newark, New Jersey, USA}

\begin{abstract}
Motivated by the recent theoretical studies on a two-dimensional (2D) chiral Hamiltonian based on the Su-Schrieffer-Heeger chains [L. Zhu, E. Prodan, and K. H. Ahn, Phys. Rev. B {\bf 99}, 041117(R) (2019)], we experimentally and computationally demonstrate that topological flat frequency bands can occur at open edges of 2D planar metamaterials and at antiphase boundary seams of ring-shaped or tubular metamaterials. Specifically, using mechanical systems made of magnetically coupled spinners, we reveal that the presence of the edge or seam bands that are flat in the entire projected reciprocal space follows the predictions based on topological winding numbers. The edge-to-edge distance sensitively controls the flatness of the edge bands and the localization of excitations, consistent with the theoretical analysis. The analog of the fractional charge state is observed. Possible realizations of flat bands in a large class of metamaterials, including photonic crystals and electronic metamaterials, are discussed. 
\end{abstract}

\maketitle

Flat energy bands have been the focus of intense research in photonic crystals, such as the Lieb lattice, due to the possibility of trapping photons, which has technological significance \cite{lieb1989two, wiersma2015trapped, vicencio2015observation, mukherjee2015observation, klembt2017polariton, leykam2018perspective, leykam2018artificial, lazarides2019compact}. They have also gained a lot of attention \cite{kariyado2019pi} following the discovery of superconductivity in twisted bilayer graphenes \cite{bistritzer2011moire, cao2018correlated, cao2018unconventional, po2018origin, yankowitz2019tuning} and the pursuit of nearly flat bands in the fractional Chern insulators \cite{tang2011high, neupert2011fractional, sun2011nearly, wang2011nearly, liu2012fractional, wang2012fractional}. Recently, there has been a theoretical proposal for flat energy bands within antiphase and twin boundaries and at open edges in a system described by a topological two-dimensional (2D) model Hamiltonian \cite{zhu2019flat}. Unlike the Lieb lattice, the flat band states occur only at edges or domain boundaries, giving a unique controllability through patterning. Unlike twisted bilayer graphenes or fractional Chern insulators, the flatness of the bands in the entire projected reciprocal space does not require tuning of parameters. In this Letter, we experimentally demonstrate the realization of the Hamiltonian and the flat bands in metamaterials, using mechanical systems made of interacting spinners \cite{apigo2018topological, qian2018topology}. By examining how the width of the edge band narrows in frequency as the edge-to-edge distance increases, we show the presence of the topological flat frequency bands at the edges. It is revealed that the size of the localized excitations at the edges correlates with the width of the edge band. The analog to electronic charge fractionalization \cite{asboth2016short} is found. We experimentally verify the presence of a midgap mode at the antiphase boundary seam of a ring-shaped spinner system, and computationally find a flat antiphase boundary seam band for a tubular system.

Systems of magnetically coupled spinners are versatile experimental platforms for various Hamiltonians \cite{apigo2018topological, qian2018topology}. In mapping between electronic tight-binding Hamiltonians and magnetically coupled spinner systems, the intersite electron hopping corresponds to the interspinner magnetic interaction controlled by the distance between the magnets. The electronic model Hamiltonians for the flat bands at open edges and twin and antiphase boundaries studied in Ref.~\cite{zhu2019flat} are based on a particular 2D extension of the Su-Schrieffer-Heeger (SSH) model \cite{su1979solitons}. Unlike other 2D SSH models \cite{xie2018second, delplace2011zak}, SSH chains with alternating intersite hopping strengths are shifted and stacked in the direction perpendicular to the chains. With only the nearest neighbor hoppings, the 2D system preserves the chiral symmetry of the one-dimensional (1D) SSH system. With the constant \emph{interchain} coupling weaker than the average \emph{intrachain} coupling, a gap opens between two bulk bands and the topology of the system is characterized by the winding number, which depends on the direction of edges or boundaries. The bulk-boundary correspondence predicts flat zero energy edge or boundary bands of bipartite states for the chiral 2D SSH system, similar to the 1D SSH system.

\begin{figure}[t]
\includegraphics[width=\linewidth]{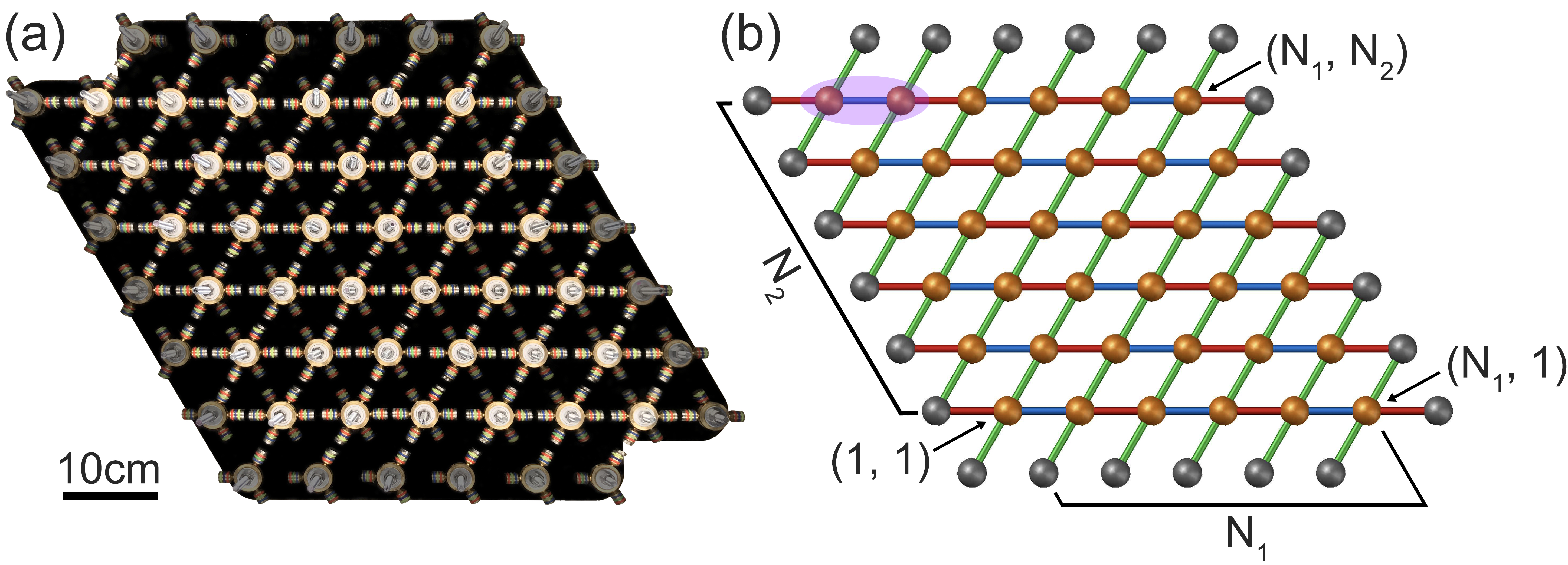}
\caption{\small (a) $6\times6$ spinner system, where rotatable spinners and magnetically coupled arms are highlighted and fixed spinners and arms without magnets are shaded. (b) Illustration of the spinner system pictured in (a), where the orange [gray] balls represent rotatable [fixed] spinners. The purple ellipse represents a unit cell. The blue [red] lines indicate the couplings within [between] the unit cells within the 1D SSH chains along the $0\degree$ direction. The green lines indicate the constant interchain coupling, which is smaller than the average intrachain coupling. Coordinates $(n_1, n_2)$ describe the position of the spinners.} 
\label{Fig:1}
\end{figure}

One of the 2D spinner systems and its schematic diagram are shown in Figs.~\ref{Fig:1}(a) and \ref{Fig:1}(b). With magnets attached to the $0\degree$, $60\degree$, $180\degree$, and $240\degree$ direction arms and the spinners arranged in quasitriangular lattices, the systems are equivalent to the electronic systems of quasisquare lattices in Ref.~\cite{zhu2019flat} with hoppings in the $0\degree$, $90\degree$, $180\degree$, and $270\degree$ directions. The systems are assembled with the edges in the $0\degree$ and $120\degree$ directions, equivalent to the $0\degree$ and $135\degree$ directions for the quasisquare systems. The spinners are indexed as $(n_1, n_2)$ with $n_1=1, ..., N_1$ and $n_2=1, ..., N_2$ (Fig.~\ref{Fig:1}(b)). The SSH chains run along the $0\degree$ direction with alternating intrachain couplings, represented by the red and the blue lines in Fig.~\ref{Fig:1}(b). The chains are coupled along the $60\degree$ direction with a constant interchain coupling, represented by the green lines. A unit cell is marked in purple. If the interchain coupling is weaker than the average intrachain coupling, and the interaction within the unit cell is weaker [stronger] than the interaction between the unit cells within the same chain, the system becomes a topological [nontopological] insulator (See Supplemental Material \cite{supp-text1}.). For chiral symmetry, fixed spinners with necessary interactions are placed around the edges, as shown in Figs.~\ref{Fig:1}(a) and \ref{Fig:1}(b). One of the spinners is driven by the interaction between a magnet on either $120\degree$ or $300\degree$ direction arm and a magnet on the actuator. The voltage from an attached accelerometer divided by the square of the frequency is used as a quantity proportional to the oscillation amplitude of the spinner. Slow motion movies are analyzed for the pattern of modes \cite{apigo2018topological, qian2018topology}. 

With parameters from Ref.~\cite{apigo2018topological}, the spectra are calculated to decide which spinners to actuate and measure, so that the bulk and edge bandwidths are well represented. Actuating and measuring at the $(N_1, 2)$ spinner [$(N_1-1, 1)$ spinner] gives the spectrum that represents the edge [bulk] bandwidth well for the topological systems. By choosing the intermagnet distances of 5.0, 8.0, and 9.0 mm for red, blue, and green lines respectively in Fig.~\ref{Fig:1}(b), we realize topological systems with the winding numbers $\nu(120\degree)=1$, $\nu(0\degree)=0$ and topological edges in $120\degree$ direction, and by choosing 8.0, 5.0, and 9.0 mm, nontopological with $\nu(120\degree)=\nu(0\degree)=0$ \cite{zhu2019flat}. The theoretical analysis for the $N_1\times N_2$ topological systems with open boundary conditions (See Supplemental Material \cite{supp-text1}.) shows that the edge states decay rapidly within a few spinner-to-spinner distances in the $n_1$ direction and the edge bandwidths for the systems with $N_2=6$ are within $5\%-15\%~(0.12-0.03~\textrm{Hz})$ from those for the large $N_2$ limit. Thus, small size systems from $4\times6$ to $12\times6$ are sufficient to reveal the trend in the edge bandwidth versus $N_1$, the distance between the topological edges.

\begin{figure}[t]
\center
\includegraphics[width=\linewidth]{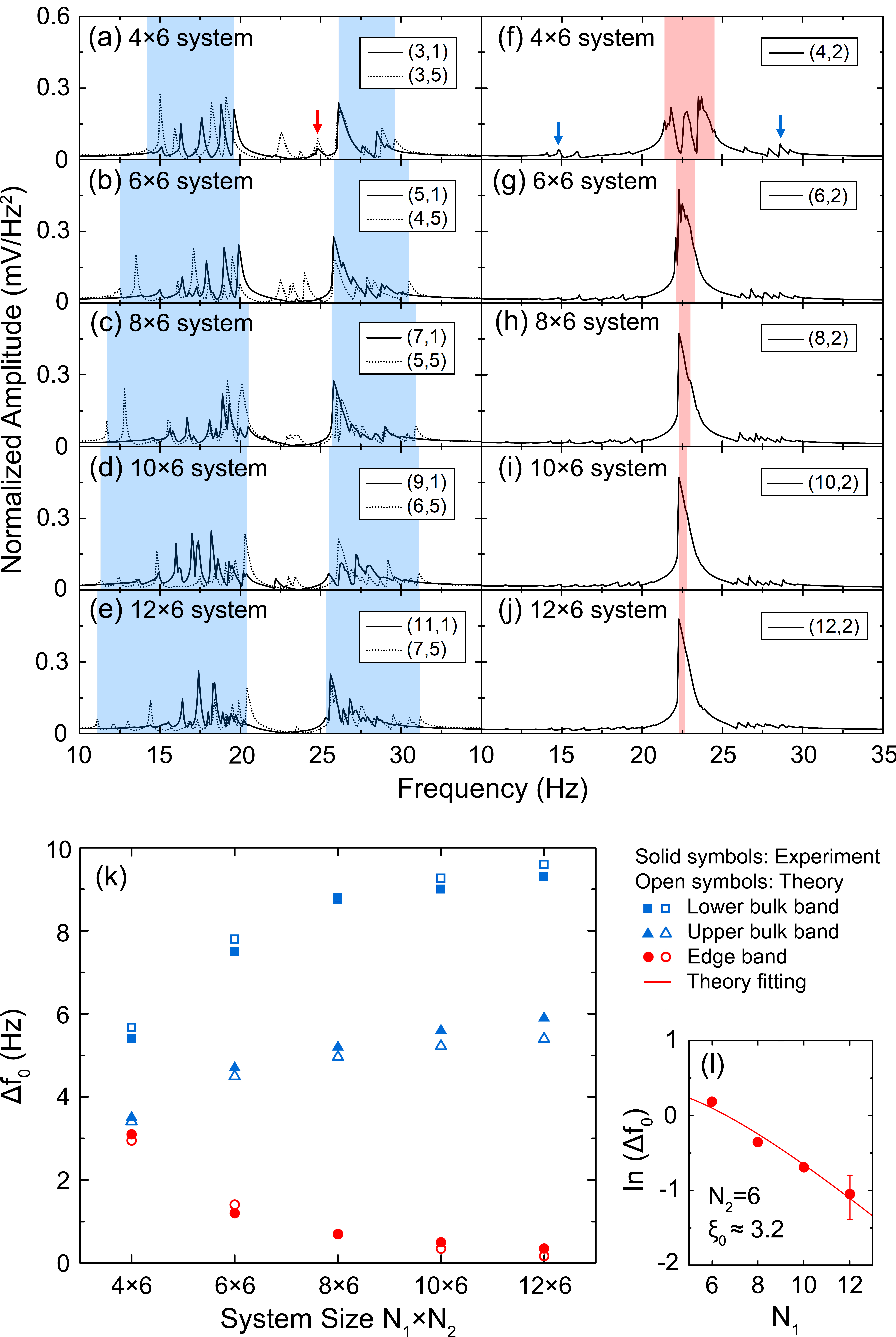}
\caption{\small Results for the topological systems. (a)-(e) Solid lines: Bulk mode spectra experimentally obtained by actuating and measuring at the $(N_1-1, 1)$ spinner for the $N_1\times6$ systems with $N_1=$ 4, 6, 8, 10, and 12, respectively. Dotted lines: Spectra obtained with the $(N_1/2+1, 5)$ spinner to reveal the top [bottom] of the upper [lower] bulk bands better. The blue areas indicate the lower and upper bulk bands. (f)-(j) Edge mode spectra experimentally obtained with the $(N_1, 2)$ spinner for the same systems as in (a)-(e), respectively. Red areas indicate the edge bands. A red arrow in (a) [blue arrows in (f)] indicates edge [bulk] modes appearing in the bulk [edge] spectra due to the short edge-to-edge distance. (k) Experimental and theoretical bulk and edge bandwidths versus the system size $N_1\times N_2$. (l) Logarithm of experimental edge bandwidth, $\ln{(\Delta f_0)}$, versus $N_1$ and fitting to the theory, resulting in the localization length at $k_2=0$, $\xi_0\approx3.2$, close to the theory value of 3.1 \cite{supp-text1, asboth2016short, footnote}.}
\label{Fig:2}
\end{figure}

Figure~\ref{Fig:2} shows the results for the topological systems. Spectra obtained with the $(N_1-1, 1)$ spinner for the $N_1\times6$ systems with $N_1=4, 6, 8, 10,$ and 12 are shown in solid lines in Figs.~\ref{Fig:2}(a)-\ref{Fig:2}(e), respectively, each of which reveals upper and lower bulk bands, marked by blue areas, and a gap in between. To reveal the modes at the top [bottom] of the upper [lower] bulk bands better, we also actuate and measure at $(N_1/2+1, 5)$ spinners, as shown in dotted lines in Figs.~\ref{Fig:2}(a)-\ref{Fig:2}(e). Figures~\ref{Fig:2}(f)-\ref{Fig:2}(j) show spectra obtained with the $(N_1, 2)$ spinner for the same systems as in Figs.~\ref{Fig:2}(a)-\ref{Fig:2}(e), respectively. Edge bands, marked by red areas, appear within the gaps of the bulk spectra. Figures~\ref{Fig:2}(a)-\ref{Fig:2}(j) show systematic changes in the bulk and edge bandwidths, which are plotted as solid symbols in Fig.~\ref{Fig:2}(k), along with the theoretical results shown as open symbols. The experimental results are in agreement with the theory and show both upper and lower bulk bandwidths increase as the edge-to-edge distance $N_1$ increases due to the finite size effect, and start to saturate around $N_1\sim8$. In contrast, the edge bandwidth from the experiments narrows rapidly as the edge-to-edge distance increases, consistent with the numerical results. Theoretical analysis for systems with a periodic boundary condition in the $n_2$ direction reveals that the edge states with the wave vector $k_2$ have bipartite patterns of zero amplitudes and exponentially decaying nonzero amplitudes with localization length $\xi(k_2)$ (See Supplemental Material \cite{supp-text1, footnote}.). This leads to the edge bandwidth $\Delta f_0=CN_1e^{-N_1/\xi_{0}}$ in the large $N_1/\xi_{0}$ limit, where $C$ is a constant and $\xi_{0}=\xi(k_2=0)$ \cite{supp-text1}. Since $N_1/\xi_{0}$ is not large for $N_1=$ 4 and the theoretical $\xi_0^{\textrm{theory}}=$ 3.1, the experimental data for $N_1=6, 8, 10$, and $12$ are used to decide $\xi_0$ experimentally as shown in Fig.~\ref{Fig:2}(l). The line represents $\ln{(\Delta f_0)}=\ln{C}+\ln{N_1}-N_1/\xi_{0}$ and shows agreement with experimental data with $\xi_0\approx3.2$, close to $\xi_0^{\textrm{theory}}$. The results in Figs.~\ref{Fig:2}(k) and \ref{Fig:2}(l) indicate that the edge band would be completely flat, as the edge-to-edge distance $N_1$ increases further, confirming the predictions from Ref.~\cite{zhu2019flat}.    

\begin{figure}[t]
\center
\includegraphics[width=\linewidth]{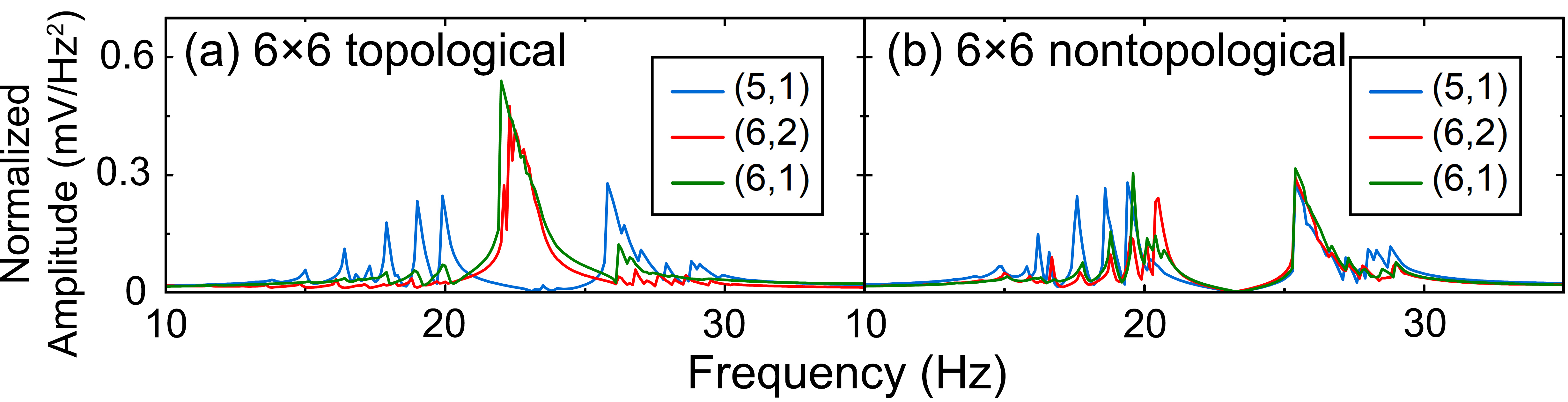}
\caption{\small Spectra experimentally obtained by actuating and measuring at the $(5, 1)$, $(6, 2)$, and $(6, 1)$ spinners for the $6\times6$ (a) topological and (b) nontopological systems. The edge band is present in the bulk band gap in (a), but absent in (b).}
\label{Fig:3}
\end{figure}

\begin{figure}[h]
\center
\includegraphics[width=\linewidth]{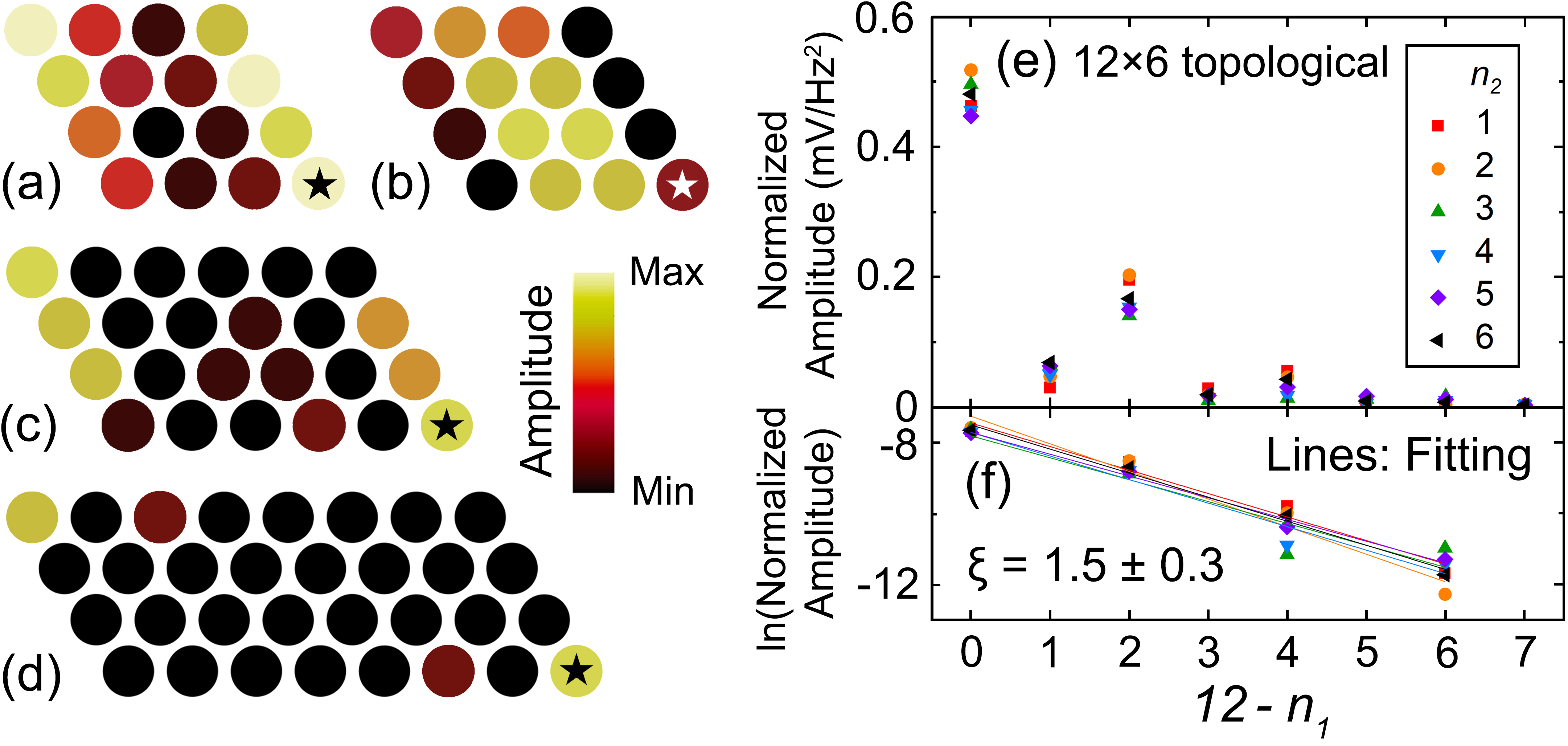}
\caption{\small (a)-(d): Patterns of modes. Colors represent the oscillation amplitudes of the spinners, estimated from slow motion movies by eyes. Stars mark actuated spinners. In (a) [(b)],  an edge [bulk] mode is revealed at 23.6 Hz [19.5 Hz] in the edge [bulk] band for the $4\times4$ topological system. In (c) [(d)], an edge mode is revealed at 23.7 Hz [24.5 Hz] for the $6\times4$ [$8\times4$] topological system. (See Supplemental Material for movies \cite{supplementalmaterial}.) (e) Normalized oscillation amplitude of the $(n_1, n_2)$ spinner versus its distance from the right edge, $12-n_1$, when a single $(12, n_2)$ spinner $(n_2=1, ..., 6)$ at the right edge is actuated for the $12\times6$ topological system. (f) Symbols: semilogarithmic plot of (e) for even $12-n_1$. Lines: linear fittings, leading to an average localization length $\xi=1.5\pm0.3$, consistent with the theoretical range of $\xi(k)$, $0.9\sim3.1$.}
\label{Fig:4}
\end{figure}

By exchanging the strong and the weak intrachain couplings, the topological systems become nontopological. The experimental results for $6\times6$ topological and nontopological systems are shown in Figs.~\ref{Fig:3}(a) and \ref{Fig:3}(b). For the topological system, the edge band is prominent in the spectra obtained from the $(6, 2)$ and $(6, 1)$ spinners, located within the gap in the spectrum from the $(5, 1)$ spinner. For the nontopological system, the edge band disappears from the gap, leaving only the bulk bands. The results show the difference between the topological and nontopological systems and the topological origin of the edge states \cite{zhu2019flat}.

For topological systems, the presence of the edge modes depends on the direction of the open edges, determined by the winding numbers $\nu(120\degree)=1$ and $\nu(0\degree)=0$. With $\nu=0$ outside the open edges, the edge modes should occur only along the $120\degree$ direction edges, not along the $0\degree$ direction. To test these predictions, we build a topological system with the same number of spinners along the $0\degree$ and $120\degree$ directions, and actuate the $(N_1, 1)$ spinner, which belongs to both $0\degree$ and $120\degree$ direction edges, at a frequency within the edge band to see along which direction the edge mode appears. We choose a small $4\times 4$ system, so that all the spinners at the topological edges show large oscillations and bulk modes could be excited by actuating the same spinner at a bulk mode frequency. The oscillation pattern at an edge mode frequency is displayed in Fig.~\ref{Fig:4}(a), in which the colors approximately represent the oscillation amplitudes of spinners, estimated from the slow motion movie by eyes (See Supplemental Material for the movie \cite{supplementalmaterial}.). The actuated spinners are marked by stars in Fig.~\ref{Fig:4}. The edge mode appears along the edges in the $120\degree$ direction, not in the $0\degree$ direction, consistent with the topological analysis. It confirms that the bands in the bulk band gap in Figs.~\ref{Fig:2}(f)-\ref{Fig:2}(j), and \ref{Fig:3}(a) are edge bands. We excite bulk modes by actuating the $(4, 1)$ spinner at a bulk band frequency, as shown in Fig.~\ref{Fig:4}(b), where the oscillations are concentrated on the central two columns along the $120\degree$ direction (For movie, see \cite{supplementalmaterial}.), confirming the bands above and below the gap in Figs.~\ref{Fig:2}(a)-\ref{Fig:2}(e) and \ref{Fig:3}(a) are the bulk bands. As the edge-to-edge distance $N_1$ increases and the edge band becomes flatter, the excitation at the edges is more localized along the edges. To verify this, the $N_1\times4$ systems with $N_1=6$ and $8$ are studied by actuating $(N_1, 1)$ spinner at edge mode frequencies. The results shown in Figs.~\ref{Fig:4}(c) and \ref{Fig:4}(d) (For movies, see \cite{supplementalmaterial}.) reveal that edge modes decay much faster along the edges compared to the $4\times4$ system in Fig.~\ref{Fig:4}(a), consistent with the enhanced localization as the edge band becomes narrower. For the $8\times4$ system shown in Fig.~\ref{Fig:4}(d), only $(1, 4)$ spinner, other than the actuated $(8, 1)$ spinner, shows a large oscillation, while the oscillations of all other spinners are much smaller, which is the analog of the fractional charge state \cite{asboth2016short}. 

\begin{figure}[t!]
\center
\includegraphics[width=\linewidth]{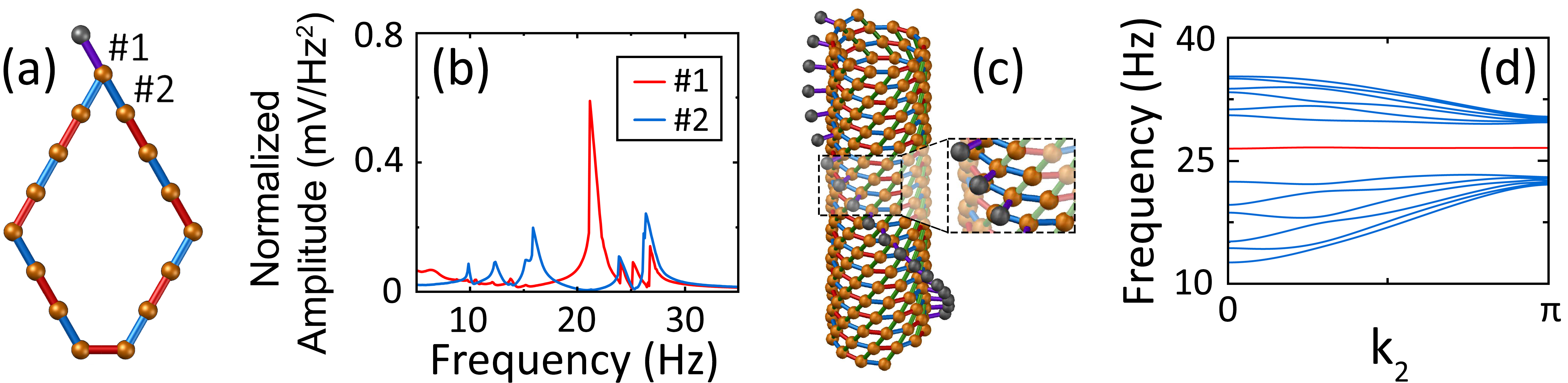}
\caption{\small  (a) Illustration of the SSH ring with fifteen spinners and a seam experimentally studied. See the caption for Fig.~\ref{Fig:1}(b). The purple line represents the extra coupling at a magnet distance of 6.5 mm for the chiral symmetry. (b) Red [Blue] line: Edge [Bulk] mode spectrum experimentally obtained by actuating and measuring at the spinner $\#$1 [$\#$2] for the SSH ring shown in (a). (See Supplemental Material for the movie of a seam mode \cite{supplementalmaterial}.) (c) Tubular model system with thirteen spinners in the azimuthal direction and an antiphase boundary seam, obtained by joining the $(1, n_2)$ and $(13, n_2+1)$ spinners of a planar system like Fig.~\ref{Fig:1}(b) with $N_1=13$. The gray balls represent fixed spinners interacting with the spinners at the seam. The magnification shows the antiphase boundary seam with the weak azimuthal couplings on both sides (the blue lines), and the couplings with the fixed spinners (the purple lines). (d) Band structure for the tubular system in (c). The blue lines represent the bulk bands and the red line the flat antiphase boundary seam band.}
\label{Fig:5}
\end{figure}

As mentioned briefly, theoretical analysis leads to the amplitude of the right edge mode with the wave vector $k_2$ vanishes for odd $n_1$, and decays as ${B}_{k_2} (n_1,n_2) = {B} e^{-(N_1-n_1)/{\xi(k_2)}}$ for even $n_1$, where $\xi(k_2)$ is the localization length \cite{supp-text1, footnote}. To verify this for the $12\times6$ system, we actuate a single $(12, n_2)$ spinner $(n_2=1, ..., 6)$ at the central frequency of the edge band and measure the amplitudes for the $(n_1, n_2)$ spinners with $12-n_1=0, ..., 7$ by accelerometers. The results for each case of $n_2$ are shown in Fig.~\ref{Fig:4}(e), which reveals that the right edge modes have much smaller amplitude for odd $n_1$ than for even $n_1$, consistent with the theory. Semilogarithmic plot for the data with even $n_1$ in Fig.~\ref{Fig:4}(f) shows exponentially decaying amplitudes, with the average localization length $\xi=1.5\pm0.3$, which is within the range of the theoretical $\xi(k_2)$ from 0.9 at $k_2=\pm\pi$ to 3.1 at $k_2=0$, reflecting that the excited edge states are combinations of edge states with different $k_2$.

Inhomogeneous systems could host flat bands at the antiphase or twin boundaries inside the bulk \cite{zhu2019flat, ahn2005electronic}. The antiphase boundary of the 1D SSH chain hosts the zero energy state in the gap, because it separates domains with different winding numbers \cite{asboth2016short, li2018schrieffer}. To force the system to have an antiphase boundary but no edges, a ring-shaped 1D SSH system with an odd number (15) of spinners is built, as shown in Fig.~\ref{Fig:5}(a), where the red and the blue lines represent the interactions between magnets separated by 5.0 and 8.0 mm, respectively. A fixed spinner, shown as a gray ball, is placed just outside the seam for the chiral symmetry. By actuating and measuring at the spinner $\#$1 [$\#$2], the antiphase boundary seam mode [bulk mode] is revealed in the spectrum, as shown in the red [blue] line in Fig.~\ref{Fig:5}(b). The seam mode peak is present in the gap, consistent with the theory \cite{zhu2019flat, asboth2016short}. The slow motion movie of the seam mode \cite{supplementalmaterial} reveals an oscillation of every other spinners in both directions from the spinner $\#$1 with decaying amplitudes, consistent with the theory \cite{asboth2016short, schomerus2013topologically} and experiments for other 1D SSH metamaterials \cite{li2018schrieffer, ota2018topological}. 

We theoretically consider tubular SSH systems with odd numbers of the spinners in the azimuthal direction, so that the antiphase boundary seams are forced to be present as shown in Fig.~\ref{Fig:5}(c). While this seam is \textit{locally} equivalent to the $120\degree$ antiphase boundary in Ref.~\cite{zhu2019flat}, the odd number of spinners in the azimuthal direction prevents us from defining a two-spinner unit cell or the topological winding number \textit{globally}, unlike the systems in Ref.~\cite{zhu2019flat}. With a periodic boundary condition along the tube axis, the frequency versus $k_2$ is calculated and shown in Fig.~\ref{Fig:5}(d), revealing an almost flat seam band (red line) inside the gap between the bulk bands (blue lines). The flat band found in this system, where the winding number is undefinable, demonstrates the robustness of the flat bands.
 
The experimental results for the spinner systems have implications for metamaterials experimentally shown to realize the 1D SSH model. It includes photonic \cite{han2019lasing, chen2019direct, xie2019visualization, xie2018second, ota2018topological, bello2019unconventional}, electronic \cite{belopolski2017novel}, acoustic \cite{li2018schrieffer}, plasmonic \cite{bleckmann2017spectral}, circuitry \cite{lee2018topolectrical, liu2019topologically, imhof2018topolectrical, serra2019observation, jiang2018experimental}, optical lattice \cite{meier2016observation}, and microwave \cite{zhu2018zak} metamaterials.  Realization of the chiral 2D SSH Hamiltonian in photonic crystals \cite{henriques2020topological} could lead to photons guided slowly along designed paths for pulse buffering \cite{leykam2018perspective, leykam2018artificial}. In electronic metamaterials, the effects of correlations would be enhanced for flat bands, resulting in strongly correlated phenomena \cite{mesaros2010electronic}. Dynamic tunability in optical lattices may be used to switch the system between topological and nontopological phases \cite{leykam2018artificial}. Extension to 3D systems would reveal the effects of polarization \cite{leykam2018perspective, leykam2018artificial}.

In summary, we have experimentally demonstrated the presence of the flat bands of the topological origin localized at open edges and antiphase boundary seams in mechanical metamaterials. The results presented here could apply to other metamaterials, potentially leading to novel phenomena and device applications.

\acknowledgments{We thank Emil Prodan for discussion. K. Q. and C. P. acknowledge support from the W. M. Keck Foundation. L. Z., K. H. A., and C. P. acknowledge the support from NJIT Faculty Seed Grant.}


\begin{thebibliography}{}

\bibitem{lieb1989two} E. H. Lieb, 
Two Theorems on the Hubbard Model, Phys. Rev. Lett. {\bf 62}, 1201 (1989).

\bibitem{wiersma2015trapped} D. Wiersma, 
Trapped in a photonic maze, Physics {\bf 8}, 55 (2015).

\bibitem{vicencio2015observation} R. A. Vicencio,  C. Cantillano, L. Morales-Inostroza, B. Real, C. Mej{\'\i}a-Cort{\'e}s, S. Weimann, A. Szameit, and M. I. Molina, 
Observation of Localized States in Lieb Photonic Lattices, Phys. Rev. Lett. {\bf 114}, 245503 (2015).

\bibitem{mukherjee2015observation} S. Mukherjee, A. Spracklen, D. Choudhury, N. Goldman, P. {\"O}hberg, E. Andersson, and R. R. Thomson, 
Observation of a Localized Flat-Band State in a Photonic Lieb Lattice, Phys. Rev. Lett. {\bf 114}, 245504 (2015).

\bibitem{klembt2017polariton} S. Klembt, T. H. Harder, O. A. Egorov, K. Winkler, H. Suchomel, J. Beierlein, M. Emmerling, C. Schneider, and S. H{\"o}fling,
Polariton Condensation in {\sl S-} and {\sl P-} Flatbands in a Two-Dimensional Lieb Lattice, Appl. Phys. Lett. {\bf 111}, 231102 (2017).

\bibitem{leykam2018perspective} D. Leykam, and S. Flach,
Perspective: Photonic flatbands, APL Photonics {\bf 3}, 070901 (2018).

\bibitem{leykam2018artificial} D. Leykam, A. Andreanov, and S. Flach,
Artificial flat band systems: From lattice models to experiments, Adv. Phys. X {\bf 3}, 1473052 (2018).

\bibitem{lazarides2019compact} N. Lazarides, and G. P. Tsironis, 
Compact localized states in engineered flat-band $\mathscr{PT}$ metamaterials, Sci. Rep. {\bf 9}, 4904 (2019).

\bibitem{kariyado2019pi} T. Kariyado, and R. Slager,
$\pi$-fluxes, semimetals, and flat bands in artificial materials, Phys. Rev. Research {\bf 1}, 032027(R) (2019).

\bibitem{bistritzer2011moire} R. Bistritzer, and A. H. MacDonald,
Moir{\'e} bands in twisted double-layer graphene, Proc. Natl. Acad. Sci. U.S.A. {\bf 108}, 12233 (2011).

\bibitem{cao2018unconventional} Y. Cao, V. Fatemi, S. Fang, K. Watanabe, T. Taniguchi, E. Kaxiras, and P. Jarillo-Herrero, 
Unconventional superconductivity in magic-angle graphene superlattices, Nature (London) {\bf 556}, 43 (2018).

\bibitem{cao2018correlated} Y. Cao, V. Fatemi, A. Demir, S. Fang, S. L. Tomarken, J. Y. Luo, J. D. Sanchez-Yamagishi, K. Watanabe, T. Taniguchi, E. Kaxiras, R. C. Ashoori, and P. Jarillo-Herrero, 
Correlated insulator behaviour at half-filling in magic-angle graphene superlattices, Nature (London) {\bf 556}, 80 (2018).

\bibitem{po2018origin} H. C. Po, L. Zou, A. Vishwanath, and T. Senthil,
Origin of Mott Insulating Behavior and Superconductivity in Twisted Bilayer Graphene, Phys. Rev. X {\bf 8}, 031089 (2018).

\bibitem{yankowitz2019tuning} M. Yankowitz, S. Chen, H. Polshyn, Y. Zhang, K. Watanabe, T. Taniguchi, D. Graf, A. F. Young, and C. R. Dean,
Tuning superconductivity in twisted bilayer graphene, Science {\bf 363}, 1059 (2019).

\bibitem{tang2011high} E. Tang, J. Mei, and X. Wen,
High-Temperature Fractional Quantum Hall States, Phys. Rev. Lett. {\bf 106}, 236802 (2011).

\bibitem{neupert2011fractional} T. Neupert, L. Santos, C. Chamon, and C. Mudry,
Fractional Quantum Hall States at Zero Magnetic Field, Phys. Rev. Lett. {\bf 106}, 236804 (2011).

\bibitem{sun2011nearly} K. Sun, Z. Gu, H. Katsura, and S. D. Sarma,
Nearly Flatbands with Nontrivial Topology, Phys. Rev. Lett. {\bf 106}, 236803 (2011).

\bibitem{wang2011nearly} F. Wang, and Y. Ran,
Nearly flat band with Chern number $C=2$ on the dice lattice, Phys. Rev. B {\bf 84}, 241103(R) (2011). 

\bibitem{liu2012fractional} Z. Liu, E. J. Bergholtz, H. Fan, and A. M. L{\"a}uchli,
Fractional Chern Insulators in Topological Flat Bands with Higher Chern Number, Phys. Rev. Lett. {\bf 109}, 186805 (2012).

\bibitem{wang2012fractional} Y. F. Wang, H. Yao, C. D. Gong, and D. N. Sheng,
Fractional quantum Hall effect in topological flat bands with Chern number two, Phys. Rev. B {\bf 86}, 201101(R) (2012).


\bibitem{zhu2019flat} L. Zhu, E. Prodan, and K. H. Ahn, 
Flat energy bands within antiphase and twin boundaries and at open edges in topological materials, Phys. Rev. B {\bf 99}, 041117(R) (2019).   

\bibitem{apigo2018topological} D. J. Apigo, K. Qian, C. Prodan, and E. Prodan,
Topological edge modes by smart patterning, Phys. Rev. Materials {\bf 2}, 124203 (2018). 

\bibitem{qian2018topology} K. Qian, D. J. Apigo, C. Prodan, Y. Barlas, and E. Prodan,
Topology of the valley-Chern effect, Phys. Rev. B {\bf 98}, 155138 (2018). 

\bibitem{asboth2016short} J. K. Asb{\'o}th, L. Oroszl{\'a}ny, and A. P{\'a}lyi,
{\sl A Short Course on Topological Insulators: Band Structure and Edge States in One and Two Dimensions} (Springer, Cham, 2016).
  
\bibitem{su1979solitons} W. P. Su, J. R. Schrieffer, and A. J. Heeger,
Solitons in Polyacetylene, Phys. Rev. Lett. {\bf 42}, 1698 (1979).

\bibitem{xie2018second} B. Y. Xie,  H. F. Wang, H. X. Wang, X. Y. Zhu, J. H. Jiang, M. H. Lu, and Y. F. Chen,
Second-order photonic topological insulator with corner states, Phys. Rev. B {\bf 98}, 205147 (2018). 

\bibitem{delplace2011zak} P. Delplace, D. Ullmo, and G. Montambaux,
Zak phase and the existence of edge states in graphene, Phys. Rev. B {\bf 84}, 195452 (2011). 

\bibitem{supp-text1} See Supplemental Material for further information on theoretical analyses and adequacy of the chosen spinner system sizes.

\bibitem{footnote} The theoretical analysis leads to the localization length of the edge states with a wave vector $k_2$, $\xi(k_2) = 4/\ln[(\beta_r^2+\beta_g^2+2\beta_r\beta_g\cos{k_2})/(\beta_b^2+\beta_g^2+2\beta_b\beta_g\cos{k_2})]$. Here, $\beta_r$, $\beta_b$, and $\beta_g$ are the parameter $\beta$ defined in Ref.~\cite{apigo2018topological} for the pairs connected by red, blue and green lines in Fig.\ref{Fig:1}(b) with values of 280.0 Hz$^2$, 106.7 Hz$^2$, and 83.4 Hz$^2$, respectively, resulting in $\xi(k_2)$ from 0.9 at $k_2=\pm\pi$ to 3.1 at $k_2=0$. See Supplemental Material for details \cite{supp-text1}.

\bibitem{supplementalmaterial} See Supplemental Material for the slow motion movies of the edge and bulk modes shown in Fig.~\ref{Fig:4} and the seam mode discussed in Fig.~\ref{Fig:5}(b) caption.


\bibitem{ahn2005electronic} K. H. Ahn, T. Lookman, A. Saxena, and A. R. Bishop,
Electronic properties of structural twin and antiphase boundaries in materials with strong electron-lattice couplings, Phys. Rev. B {\bf 71}, 212102 (2005).   

\bibitem{li2018schrieffer} X. Li, Y. Meng, X. Wu, S. Yan, Y. Huang, S. Wang, and W. Wen,
Su-Schrieffer-Heeger model inspired acoustic interface states and edge states, Appl. Phys. Lett. {\bf 113}, 203501 (2018).

\bibitem{schomerus2013topologically} H. Schomerus,
Topologically protected midgap states in complex photonic lattices, Opt. Lett. {\bf 38}, 1912 (2013).

\bibitem{ota2018topological} Y. Ota, R. Katsumi, K. Watanabe, S. Iwamoto, and Y. Arakawa,
Topological photonic crystal nanocavity laser, Commun. Phys. {\bf 1}, 86 (2018).   

\bibitem{han2019lasing} C. Han, M. Lee, S. Callard, C. Seassal, and H. Jeon,
Lasing at topological edge states in a photonic crystal L3 nanocavity dimer array, Light Sci. Appl. {\bf 8}, 40 (2019).   

\bibitem{bello2019unconventional} M. Bello, G. Platero, J. I. Cirac, and A. Gonz{\'a}lez-Tudela,
Unconventional quantum optics in topological waveguide QED, Sci. Adv. {\bf 5}, eaaw0297 (2019).   

\bibitem{chen2019direct} X. D. Chen, W. M. Deng, F. L. Shi, F. L. Zhao, M. Chen, and J. W. Dong,
Direct Observation of Corner States in Second-Order Topological Photonic Crystal Slabs, Phys. Rev. Lett. {\bf 122}, 233902 (2019).

\bibitem{xie2019visualization} B. Y. Xie, G. X. Su, H. F. Wang, H. Su, X. P. Shen, P. Zhan, M. H. Lu, Z. L. Wang, and Y. F. Chen,
Visualization of Higher-Order Topological Insulating Phases in Two-Dimensional Dielectric Photonic Crystals, Phys. Rev. Lett. {\bf 122}, 233903 (2019).

\bibitem{belopolski2017novel} I. Belopolski \emph{et al.}, 
A novel artificial condensed matter lattice and a new platform for one-dimensional topological phases, Sci. Adv. {\bf 3}, e1501692 (2017).

\bibitem{bleckmann2017spectral} F. Bleckmann, Z. Cherpakova, S. Linden, and A. Alberti
Spectral imaging of topological edge states in plasmonic waveguide arrays, Phys. Rev. B {\bf 96}, 045417 (2017).  

\bibitem{lee2018topolectrical} C. H. Lee, S. Imhof, C. Berger, F. Bayer, J. Brehm, L. W. Molenkamp, T. Kiessling, and R. Thomale,
Topolectrical circuits, Commun. Phys. {\bf 1}, 39 (2018). 

\bibitem{imhof2018topolectrical} S. Imhof, C. Berger, F. Bayer, J. Brehm, L. W. Molenkamp, T. Kiessling, F. Schindler, C. H. Lee, M. Greiter, T. Neupert, and R. Thomale, 
Topolectrical-circuit realization of topological corner modes, Nat. Phys. {\bf 14}, 925 (2018).

\bibitem{liu2019topologically} S. Liu, W. Gao, Q. Zhang, S. Ma, L. Zhang, C. Liu, Y. J. Xiang, T. J. Cui, and S. Zhang,
Topologically protected edge state in two-dimensional Su-Schrieffer-Heeger circuit, Research {\bf 2019}, 8609875 (2019).

\bibitem{serra2019observation} M. Serra-Garcia, R. S{\"u}sstrunk, and S. D. Huber,
Observation of quadrupole transitions and edge mode topology in an LC circuit network, Phys. Rev. B {\bf 99}, 020304(R) (2019).

\bibitem{jiang2018experimental} J. Jiang, Z. Guo, Y. Ding, Y. Sun, Y. Li, H. Jiang, and H. Chen,
Experimental demonstration of the robust edge states in a split-ring-resonator chain, Opt. Express {\bf 26}, 12891 (2018).

\bibitem{meier2016observation} E. J. Meier, F. A. An, and B. Gadway,
Observation of the topological soliton state in the Su-Schrieffer-Heeger model, Nat. Commun. {\bf 7}, 13986 (2016).

\bibitem{zhu2018zak} W. Zhu, Y. Ding, J. Ren, Y. Sun, Y. Li, H. Jiang, and H. Chen,
Zak phase and band inversion in dimerized one-dimensional locally resonant metamaterials, Phys. Rev. B {\bf 97}, 195307 (2018).

\bibitem{henriques2020topological} J. C. G. Henriques, T. G. Rappoport, Y. V. Bludov, M. I. Vasilevskiy, and N. M. R. Peres,
Topological photonic Tamm states and the Su-Schrieffer-Heeger model, Phys. Rev. A {\bf 101}, 043811 (2020).

\bibitem{mesaros2010electronic} A. Mesaros, S. Papanikolaou, C. F. J. Flipse, D. Sadri, and J. Zaanen,
Electronic states of graphene grain boundaries, Phys. Rev. B {\bf 82}, 205119 (2010).  


\end{thebibliography}
\end{document}